\documentclass{iopart}
\usepackage{graphicx}  
\usepackage{latexsym}
\usepackage{amssymb}

\jl{6}        
\eqnobysec    

\def\beq{\begin{equation}}
\def\eeq{\end{equation}}
\def\rmd{{\rm d}} 

\def\version{\today}

\begin{document}

\begin{flushright}
Current version: \version 
\end{flushright}

\title[de Sitter spacetime: effects of metric perturbations on geodesics]
{de Sitter spacetime: effects of metric perturbations on geodesic motion}

\author{
Donato Bini$^* {}^\S{}^\dag$,
Giampiero Esposito$^\P$
and
Andrea Geralico${}^\S{}^\ddag$
}

\address{
  ${}^*$\
Istituto per le Applicazioni del Calcolo ``M. Picone,'' CNR, 
00185 Rome, Italy
}

\address{
  ${}^\S$\
  ICRA,
  University of Rome ``La Sapienza,'' 00185 Rome, Italy
}

\address{
  ${}^\dag$\
  INFN, Sezione di Firenze, 00185
  Sesto Fiorentino (FI), Italy
}

\address{
${}^\P$\
  INFN, Sezione di Napoli, 
Complesso Universitario di Monte S. Angelo,
Via Cintia, Edificio 6, 80126 Napoli, Italy
}

\address{
  $^\ddag$
  Physics Department,
  University of Rome ``La Sapienza,'' 00185 Rome, Italy
}

\begin{abstract}
Gravitational perturbations of the de Sitter spacetime are investigated using the Regge--Wheeler formalism.
The set of perturbation equations is reduced to a single second order differential equation of the Heun-type for both electric and magnetic multipoles.
The solution so obtained is used to study the deviation from an initially radial geodesic due to the perturbation.
The spectral properties of the perturbed metric are also analyzed. 
Finally, gauge- and tetrad-invariant first-order massless perturbations of any spin are explored following the approach of Teukolsky.
The existence of closed-form, i.e. Liouvillian, solutions to the radial part of the Teukolsky master equation is discussed.
\end{abstract}

\pacno{04.20.Cv, 04.30.Nk}

\section{Introduction}

de Sitter spacetime is extensively investigated in the literature from different points of view.
In fact, it is particularly relevant for inflation \cite{Guth1981} and hence the physics of the very early universe.
Furthermore, there are many recent experimental evidences and astronomical observations which indicate that the
cosmological constant in the universe is positive leading to an accelerating expansion (see, e.g., Ref. \cite{bass} and references therein).
The pioneering work of Gibbons and Hawking \cite{gibbons} extending to cosmological horizons the concepts of thermodinamics and particle creation previously introduced for black hole event horizons opened the way to the study of the properties of quantum field theory on de Sitter space, especially the temperature and entropy (see, e.g., Refs. \cite{lohiya1,lohiya2,biswas,suzuki}), and on asymptotically de Sitter spaces containing black holes.
In this respect, it proved useful to express the de Sitter metric in static form (see, e.g., Ref. \cite{ellis}).

We are interested here in studying metric perturbations in the static region of the de Sitter spacetime between the origin and the cosmological horizon.
We will then use the Regge--Wheeler \cite{ReggeW} formalism to decompose the perturbations in tensor harmonics and Fourier transform it with respect to time, as customary.
We will account for perturbations of both parity, and investigate the features of geodesic motion in the perturbed gravitational field.
In particular, we will consider an initially radial geodesic in the unperturbed spacetime and study how the motion deviates from maintaining purely radial as a result of the perturbation. 
Furthermore, we will also consider gauge- and tetrad-invariant first-order massless perturbations of any spin following the approach of Teukolsky \cite{Teuk}, and discuss the existence of closed-form, i.e. Liouvillian, solutions to the radial equation.
Note that our results are consistent with those of Refs. \cite{guven} and \cite{khanal} concerning the perturbations of the Schwarzschild-de Sitter spacetime, which however focused on the stability of the de Sitter spacetime in the presence of the black hole and on the scattering properties of both the event and cosmological horizons. 
 
Section 2 summarizes what is known about radial geodesics in de Sitter
spacetime, while metric perturbations for both electric and magnetic 
multipoles are discussed in section 3. The equations for perturbations
to radial geodesics are integrated in section 4, while section 5 contains the derivation of Liouvillian solutions to the radial part of the Teukolsky master equation for massless perturbations of any spin. 
Concluding remarks are presented in
section 6, while relevant details are given in the appendices.

\section{Radial geodesics in de Sitter spacetime}

Consider de Sitter spacetime, whose line element written in  
spherical-like coordinates is given by
\beq 
\label{metric}
\rmd  s^2 = N^2\rmd t^2 - N^{-2} \rmd r^2 
- r^2 (\rmd \theta^2 +\sin^2 \theta \rmd \phi^2),
\eeq
where $N=(1-H^2r^2)^{1/2}$ denotes the \lq\lq lapse" function. 
The spacetime region which can be accessed is the ball of radius 
$1/H$ and center at the origin.
Timelike radial geodesics are characterized by their $4$-velocity
\beq
\label{rad_geo}
U=\Gamma (\partial_t +\zeta \partial_r)
\eeq
with 
\beq
\Gamma=\frac{\rmd t}{\rmd \tau}=\frac{E}{N^2},\qquad 
\frac{\rmd r}{\rmd \tau}=\Gamma \zeta=\pm \sqrt{E^2-N^2},
\eeq
so that $\tau$ denotes the proper time and 
\beq
\zeta=\pm \frac{N^2}{E}\sqrt{E^2-N^2};
\eeq
$E$ is the Killing constant associated with the stationarity of the metric.
It is convenient to introduce an orthonormal frame adapted to the 
static observer family (at rest with respect to the coordinates)
with basis vector fields
\beq
\fl\quad
e_{\hat 0}=N^{-1}\partial_t,\quad e_{\hat r}
=N\partial_r,\quad e_{\hat \theta}=1/r\partial_\theta,
\quad e_{\hat \phi}=1/(r\sin\theta)\partial_\phi,
\eeq
with dual basis of 1-forms
\beq
\fl\quad
\omega^{\hat 0}=N dt,\quad \omega^{\hat r}=-N^{-1}dr,
\quad \omega^{\hat \theta}=-rd\theta,\quad \omega^{\hat \phi}
=-r\sin\theta d\phi.
\eeq
With respect to this frame the geodesic $4$-velocity vector field can 
be written as
\beq
U=\gamma (e_{\hat 0}+\nu e_{\hat r}),
\eeq
where
\beq
\gamma=\Gamma N,\quad \nu=N^{-2}\zeta.
\eeq
The parametric equations of the geodesics corresponding to initial values 
\beq
\label{iniconds}
t(0)=0\,,\quad r(0)=0,\quad \theta(0)=0,\quad \phi(0)=0
\eeq
can be written explicitly as follows.
\begin{enumerate}
\item $E^2>1$.

In this case it is convenient to use the notation $E=\cosh \alpha$.
We have then
\begin{eqnarray}
\fl
t(\tau)&=& \frac{1}{H}\left[ {\rm arctanh}\left(\frac{\tanh 
(\frac12 H\tau)+\sinh \alpha}{\cosh \alpha}\right) 
+{\rm arctanh}\left(\frac{\tanh (\frac12 H\tau)-\sinh \alpha}
{\cosh \alpha}\right)
\right]\nonumber \\
\fl
r(\tau)&=& \frac{\sinh \alpha}{H}\sinh (H\tau)
\end{eqnarray}

\item $E^2=1$.

In this case the initial conditions (\ref{iniconds}) imply the 
trivial solution of a particle at rest at the origin
\beq
\label{solEeq1}
r=0,\quad t=\tau.
\eeq

\item $E^2<1$.

In this case it is convenient to use the notation $E=\cos  \alpha$. 
However $r(0)$ cannot be made to vanish but it should be always 
greater than a certain value  $0<r_*=\sin \alpha/H<1/H$.
We have  then
\begin{eqnarray}
\fl
t(\tau)&=& \frac{1}{H}\left\{ 
{\rm arctanh }\left[ \frac{1+\sin\alpha}{\cos\alpha} \tanh 
\left(\frac12 H\tau \right)\right]
+ {\rm arctanh }\left[ \frac{1-\sin\alpha}{\cos\alpha} \tanh 
\left(\frac12 H\tau \right)\right]
\right\}\nonumber \\
\fl
r(\tau)&=& \frac{\sin  \alpha}{H} \cosh  (H\tau) \equiv r_*  
\cosh  (H\tau).
\end{eqnarray}
\end{enumerate}
With a slight modification one can obtain the solution corresponding to 
more general initial data $x^\alpha(0)=x^\alpha_0$.
For instance, for $r(0)\not=0$ the solution (\ref{solEeq1}) for $E^2=1$ becomes
\beq
r=r_0e^{\pm H\tau}, \qquad
t=\tau\mp\frac1H\ln\frac{N}{N_0}.
\eeq 

For a later use we also introduce the Newman--Penrose 
\cite{NP} frame
\beq
\fl\qquad
l=\frac{1}{N^2}\partial_t+\partial_r,\quad 
n=\frac{N^2}{2}\left(\frac{1}{N^2}\partial_t-\partial_r\right),\quad 
m=\frac{1}{\sqrt{2}r}\left(\partial_\theta+\frac{{\rm i}}
{\sin\theta}\partial_\phi\right).
\eeq
The non-vanishing spin coefficients are 
\beq\fl\qquad
\rho=-\frac{1}{r},\quad
\mu=-\frac{N^2}{2r},\quad
\alpha=-\beta=-\frac{\sqrt{2}}{4r}\cot\theta,\quad
\gamma=-\frac{H^2r}{2}.
\eeq
The Weyl scalars are all identically zero.

\section{Metric perturbations}

Let us consider the perturbations 
\beq
\tilde g_{\mu \nu }=g_{\mu \nu } + h_{\mu \nu}
\eeq
of de Sitter spacetime.
Since the background metric (\ref{metric}) is static and 
spherically symmetric, we decompose the metric perturbation 
$h_{\mu \nu }$ in tensor harmonics and Fourier transform it with 
respect to time, as customary. 
We will use the Regge-Wheeler \cite{ReggeW} gauge to simplify the 
form of the perturbation.

\subsection{Electric multipoles}

The metric perturbation $h_{\mu \nu }$, for the electric multipoles, 
is given by
\begin{eqnarray}
\label{elepert}
 ||h_{\mu \nu }||=\left[ 
\begin {array}{cccc} 
N^2H_0&H_1&0&0
\\\noalign{\medskip}\rm{sym}&\displaystyle\frac{1}{N^2}H_2&0&0
\\\noalign{\medskip}\rm{sym}&\rm{sym}&{r}^{2}K&0\\\noalign{\medskip}
\rm{sym}&\rm{sym}
&\rm{sym}&{r}^{2} \sin ^{2}\theta K
\end {array} 
\right]{\rm e}^{-{\rm i}\omega t}Y_{l0} ,
\end{eqnarray}
where the symbol ``sym'' inherited from \cite{ReggeW} 
indicates that the missing components 
of $h_{\mu \nu }$ should be found from the symmetry 
$h_{\mu \nu }=h_{\nu \mu }$ and the functions $Y_{l0}$ are 
normalized spherical harmonics (see e.g. \cite{jackson}) with 
azimuthal index $m=0$, defined by 
\begin{eqnarray}
\label{ypsilonl0}
Y_{l0}=\frac12\sqrt{{\frac {2l+1}{\pi }}}P_l(\cos\theta) .
\end{eqnarray}

For $l\geq2$, the system of radial equations we have to solve is 
the following (here $L \equiv l(l+1)$):
\begin{eqnarray}
\label{eqsele}  
0&=&H_1{}'+\frac{{\rm i} \omega}{N^2}(W+K)-\frac{2rH^2}{N^2}H_1 , 
\nonumber\\
0&=&K{}'-\frac{W}{r}-\frac{{\rm i}L}{2\omega r^2}H_1+\frac{K}{rN^2} , 
\end{eqnarray}
since 
\beq
H_0=H_2\equiv W , \qquad
\eeq
together with the algebraic relation
\beq
\fl\qquad
\label{constr}
0=\frac{(L-2)}{r}W-\frac{{\rm i}}{\omega}(H^2L+2\omega^2)H_1
+\frac{2}{rN^2}\left(1+r^2\omega^2-N^2\frac{L}{2}\right)K .
\eeq

Let us introduce the dimensionless variables $x \equiv Hr$, 
$\Omega \equiv \omega/H$, and set $H_{1} \equiv {\rm i} \tilde H_1$.
Solving for $W$ in the constraint equation (\ref{constr}) and 
substituting into the system (\ref{eqsele}) we get (from now on a 
prime denotes differentiation with respect to $x$)
\begin{eqnarray}
\label{eqsele2a}  
\tilde H_1{}'&=&\frac{(2\Omega^2+3L-4)x}{(1-x^2)(L-2)}\tilde H_1
+\frac{2\Omega[(\Omega^2+L-1)x^2+2-L]}{(1-x^2)^2(L-2)}K , \\
\label{eqsele2b}
K{}'&=&-\frac{[2x^2(L+2\Omega^2)+L(L-2)]}{2(L-2)\Omega x^2}
\tilde H_1-\frac{x(L+2\Omega^2)}{(1-x^2)(L-2)}K , 
\end{eqnarray}
which can be solved in terms of Heun's functions (see Appendix C).

Figure \ref{fig:1} shows the behaviour of the metric perturbation 
functions $\tilde H_1$, $K$ and $W$ in terms of the rescaled radial 
variable $x$ for $l=2$ and different values of $\Omega$.
The perturbative regime is generally lost very quickly.
High frequency modes instead oscillate with small amplitudes over 
a wide range of the radial coordinate.
The oscillations become very high when approaching the de Sitter 
horizon in the case of $\tilde H_1$ and $W$.
The amplitude remains instead practically constant in the case of $K$.
In fact, in the high frequency limit $\Omega\gg1$ one finds 
\beq
K{}''+\frac{2x}{1-x^2}K{}'+\frac{\Omega^2}{(1-x^2)^2}K=0,
\eeq
or equivalently
\beq
\frac{d^2K }{dz^2}+\Omega^{2}K=0,
\eeq
having defined $z = {\rm arctanh}\,x$.
The solution is then 
\beq
K=c_1\sin(\Omega\,{\rm arctanh}\,x)+c_2\cos(\Omega\,{\rm arctanh}\,x),
\eeq
where $c_1$ and $c_2$ are arbitrary constants.


\begin{figure} 
\typeout{*** EPS figure 1}
\begin{center}
$\begin{array}{cc}
\includegraphics[scale=0.3]{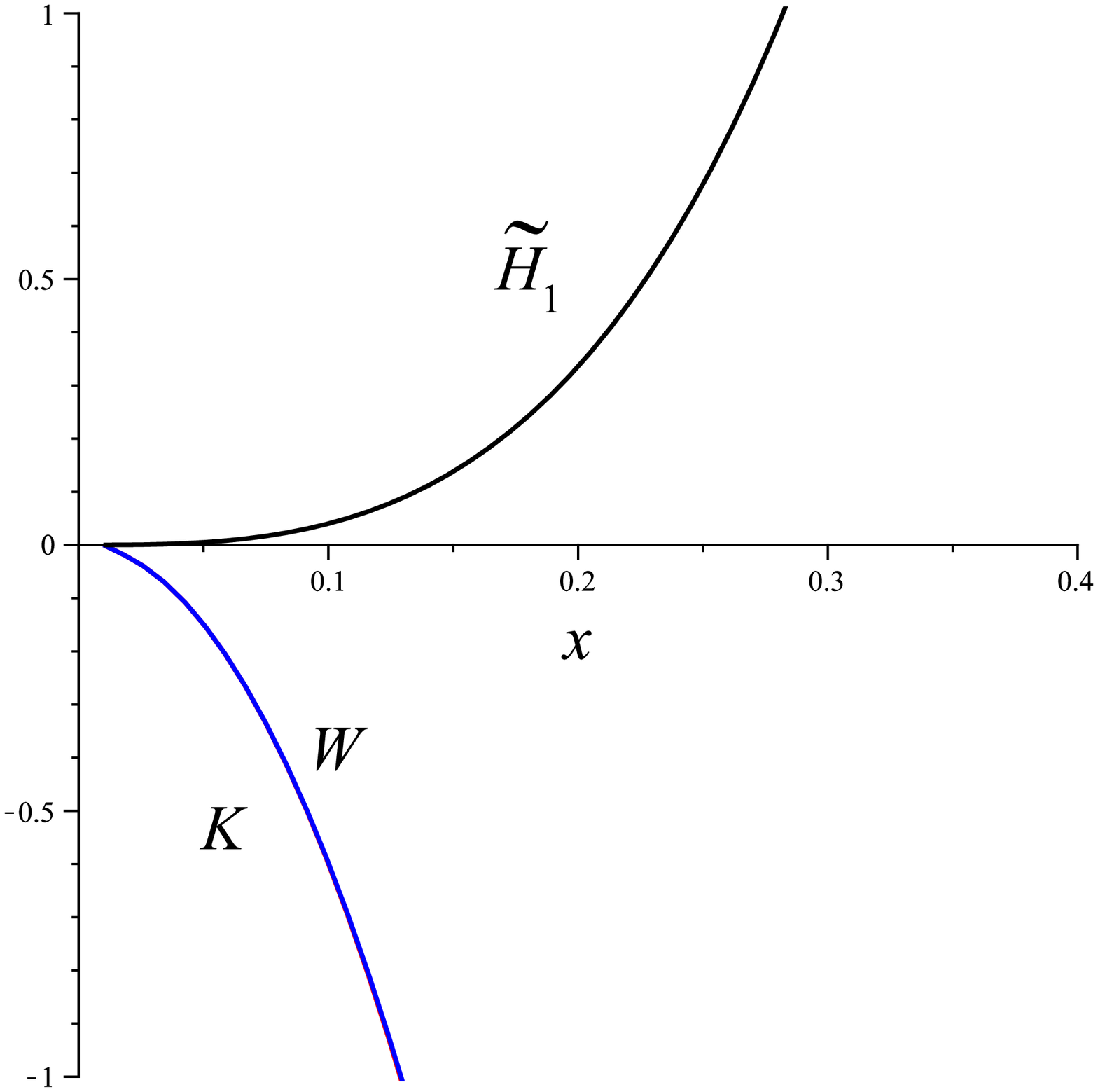}&\quad
\includegraphics[scale=0.3]{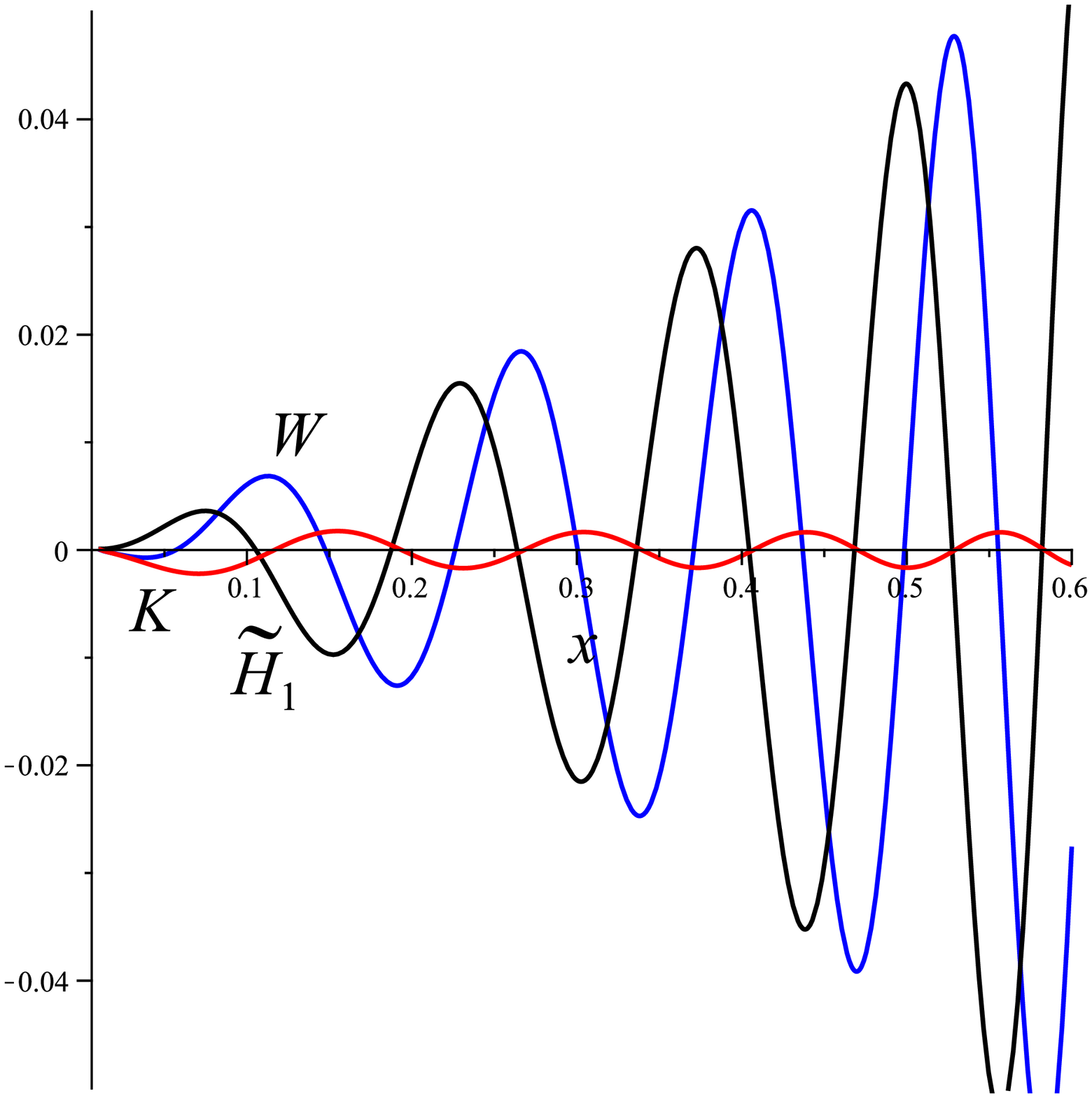}\\[.4cm]
\quad\mbox{(a)}\quad &\quad \mbox{(b)}
\end{array}$\\
\end{center}
\caption{The behaviour of the metric perturbation functions  
$\tilde H_1$, $K$ and $W$ 
in the case of electric multipoles is shown in terms 
of the rescaled radial variable $x$ for $l=2$ and different values 
of $\Omega$. Fig. (a) corresponds to the choice $\Omega=1$ 
and the initial conditions are $\tilde H_1(x_0)=10^{-4}=K(x_0)$, 
with $x_0=10^{-2}$, so that $W(x_0)\approx10^{-4}$.
Fig. (b) corresponds to $\Omega=40$ and initial conditions 
$\tilde H_1(x_0)=10^{-4}=K(x_0)$, with $x_0=10^{-2}$, so that 
$W(x_0)\approx7\times10^{-5}$.
Larger values of $l$ correspond to a further growth of the 
perturbation functions.}
\label{fig:1}
\end{figure}

A real solution for the metric can be obtained by considering $W$ 
and $K$ and $\tilde H_1$ as real. In this case the non-vanishing 
metric components are
\begin{eqnarray}
g_{tt}&=& N^2 (1+WY_{l0} \cos \omega t),
\nonumber \\
g_{tr}&=& \tilde H_1Y_{l0} \sin \omega t,
\nonumber \\
g_{rr}&=& -N^{-2} (1-WY_{l0} \cos \omega t),
\nonumber \\
g_{\theta\theta}&=& -r^2 (1-KY_{l0} \cos \omega t),
\nonumber \\
g_{\phi\phi}&=& -r^2\sin^2\theta (1-KY_{l0} \cos \omega t),
\end{eqnarray}
so that
\begin{eqnarray}
ds^2&=&N^2 (1+WY_{l0} \cos \omega t) \left(dt-\tilde H_1N^{-2}Y_{l0} 
\sin\omega t dr\right)^{2}
\nonumber \\
&&-N^{-2} (1-WY_{l0} \cos \omega t) dr^2\nonumber \\
&& -r^2 (1-KY_{l0} \cos \omega t)(d\theta^2+\sin^2\theta d\phi^2),
\end{eqnarray}
to first order in the perturbation quantities.
A natural orthonormal frame associated with this form of the metric is then
\begin{eqnarray}
\Omega^{\hat 0}&=& N  \left(1+\frac12WY_{l0} \cos \omega t\right)
\left(dt+\tilde H_1N^{-2}Y_{l0} \sin\omega t dr\right)
\nonumber \\
&=& N  \left(1+\frac12WY_{l0} \cos \omega t\right)dt+\tilde H_1N^{-1}Y_{l0} 
\sin\omega tdr,
\nonumber \\
\Omega^{\hat r}&=& -N^{-1} \left(1-\frac12WY_{l0} \cos 
\omega t\right) dr, 
\nonumber \\
\Omega^{\hat \theta}&=& -r\left(1-\frac12KY_{l0} \cos 
\omega t\right)d\theta, 
\nonumber \\
\Omega^{\hat \phi}&=& -r \sin \theta \left(1-\frac12KY_{l0} \cos 
\omega t\right)  d\phi .
\end{eqnarray}
One can then introduce a Newman--Penrose 
\cite{NP} frame in a standard way \cite{chandra}
\beq
\label{NPframe}
\fl
l=(\Omega^{\hat 0}+\Omega^{\hat r})/\sqrt{2},\quad n
=(\Omega^{\hat 0}-\Omega^{\hat r})/\sqrt{2},\quad m
=(\Omega^{\hat \theta}+{\rm i} \Omega^{\hat \phi})/\sqrt{2},
\eeq
and conveniently rotate it according to $l\to l\sqrt{2}/N$, 
$n\to nN/\sqrt{2}$. We list here the non-vanishing first-order 
spin coefficients and Weyl scalars (hereafter
$C \equiv \cos \omega t$ and $ S \equiv \sin \omega t$): 
\begin{eqnarray}
\rho&=&-\frac{1}{r}+\frac{1}{2r}(rK'-W)Y_{l0}C-\frac{\omega}
{2N^2}KY_{l0}S,
\nonumber\\
\mu&=&-\frac{N^2}{2r}+\frac{N^2}{4r}(rK'-W)Y_{l0}C+\frac{\omega}{4}
KY_{l0}S,
\nonumber\\
\alpha&=&-\frac{\sqrt{2}}{4r}\cot\theta+\frac{\sqrt{2}}{8r}K
\left(\frac{dY_{l0}}{d\theta}-\cot\theta Y_{l0}\right)C
-\frac{\sqrt{2}}{8r}\tilde H_1 Y_{l0}S,
\nonumber\\
\beta&=&-\alpha-\frac{\sqrt{2}}{4r}\tilde H_1\frac{dY_{l0}}
{d\theta}S,
\nonumber\\
\gamma&=&-\frac{H^2r}{2}-\frac18(-N^2W'+2\omega\tilde H_1
+2rH^2W)Y_{l0}C-\frac{\omega}{8}WY_{l0}S,
\nonumber\\
\nu&=&\frac{\sqrt{2}N^2}{8r}(WC-\tilde H_1S)\frac{dY_{l0}}{d\theta},
\nonumber\\
\epsilon&=&\frac{Y_{l0}}{4N^2}[(N^2W'-2\omega\tilde H_1)C
+\omega WS],
\nonumber\\
\kappa&=&-\frac{\sqrt{2}}{2N^2r}(WC+\tilde H_1S)\frac{dY_{l0}}{d\theta},
\end{eqnarray}
and
\begin{eqnarray}
\label{psiele}
\psi_0&=&-\frac{1}{2N^2r^2}(WC+\tilde H_1S){}_2Y^0{}_l,
\nonumber\\
\psi_1&=&\frac{\sqrt{2}}{4N^2r^2}\left[\frac{(\tilde H_1LN^2
+2r\omega K)}{2r\omega}C+(\tilde H_1N^2+r\omega K)S\right]
\frac{dY_{l0}}{d\theta},
\nonumber\\
\psi_2&=&\frac{L}{4\omega r^3}(\tilde H_1N^2+r\omega K)Y_{l0}C,
\nonumber\\
\psi_3&=&-\frac{\sqrt{2}}{8r^2}\left[\frac{(\tilde H_1LN^2
+2r\omega K)}{2r\omega}C-(\tilde H_1N^2+r\omega K)S\right]
\frac{dY_{l0}}{d\theta},
\nonumber\\
\psi_4&=&-\frac{N^2}{8r^2}(WC-\tilde H_1S){}_2Y^0{}_l,
\end{eqnarray}
where
\beq
{}_2Y^0{}_l=2\frac{dY_{l0}}{d\theta}\cot\theta +LY_{l0}.
\eeq
Note that the first-order Ricci scalar coincides with the zeroth-order one.

\begin{figure} 
\typeout{*** EPS figure S_ele}
\begin{center}
\includegraphics[scale=0.3]{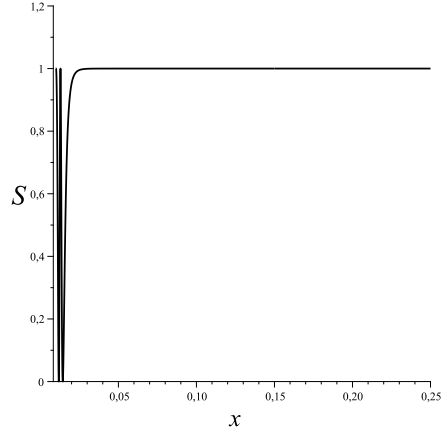}
\end{center}
\caption{The behaviour of the spectral index $\mathcal{S}$ in the case of electric multipoles is shown in terms 
of the rescaled radial variable $x$ for fixed values of $t=\pi/4$, $\theta=\pi/2$, $l=2$ and $\Omega=1$. 
The choice of initial conditions is the same as in Fig. \ref{fig:1} (a).
The variation of radial coordinate is limited to the range where the perturbative regime holds.
}
\label{fig:S_ele}
\end{figure}

It is useful to introduce the speciality index $\mathcal{S}$, which is a suitable combination of Weyl scalars defined by
\beq
\label{specdef}
\mathcal{S}=\frac{27J^2}{I^3}
=27\frac{(\psi_0\psi_2\psi_4-\psi_2^3)^2}{(\psi_0\psi_4+3\psi_2^2)^3},
\eeq
and has the value 1 for algebraically special spacetimes. 
Note that the imaginary part of $\mathcal{S}$ turns out to be identically vanishing, so that it is a purely real quantity.
Its behaviour as a function of the rescaled radial coordinate $x$ for fixed values of $t$ and $\theta$ is shown in Fig. \ref{fig:S_ele}.


\subsection{Magnetic multipoles}

The metric perturbation $h_{\mu \nu }$, for magnetic multipoles, is given by
\begin{eqnarray}
\label{magnpert}
 ||h_{\mu \nu }||=\left[ 
\begin {array}{cccc} 
0&0&0&h_0
\\\noalign{\medskip}\rm{sym}&0&0&h_1
\\\noalign{\medskip} \rm{sym} &\rm{sym} &0&0\
\\\noalign{\medskip} \rm{sym} & \rm{sym} & \rm{sym} &0
\end {array} 
\right]{\rm e}^{-{\rm i}\omega t}\sin\theta\frac{dY_{l0}}{d \theta}.
\end{eqnarray}

For $l\geq2$, the system of radial equations we have to solve is 
\begin{eqnarray}
\label{eqsmagn}  
0&=&h_0{}'-\frac{2h_0}{r}+{\rm i}h_1\left[\omega-\frac{N^2}{r^2}
\frac{(L-2)}{\omega}\right] , 
\nonumber\\
0&=&h_1{}'-\frac{2rH^2}{N^2}h_1-\frac{{\rm i}\omega}{N^4}h_{0} ,
\end{eqnarray}
where again $L \equiv l(l+1)$.

By introducing the dimensionless variables 
$x \equiv Hr$, $\Omega \equiv \omega/H$ 
as above and setting $h_{0} \equiv i\tilde h_0$, 
the system (\ref{eqsmagn}) becomes
\begin{eqnarray}
\label{eqsmagn2a}  
\tilde h_0{}'&=&\frac{2}{x}\tilde h_0-\frac{[(\Omega^2+L-2)x^2
+2-L]}{\Omega x^2}h_{1} , \\
\label{eqsmagn2b}
h_1{}'&=&\frac{\Omega}{(1-x^2)^2}\tilde h_0+\frac{2x}{1-x^2}h_{1} , 
\end{eqnarray}
which can be solved in terms of Heun's functions (see appendix C).

Figure \ref{fig:2} shows the behaviour of the metric perturbation 
functions $\tilde h_0$ and $h_1$ in terms of the rescaled radial 
variable $x$ for $l=2$ and different values of $\Omega$.
The same considerations holding for electric multipoles apply also 
in this case.


\begin{figure} 
\typeout{*** EPS figure 2}
\begin{center}
$\begin{array}{cc}
\includegraphics[scale=0.3]{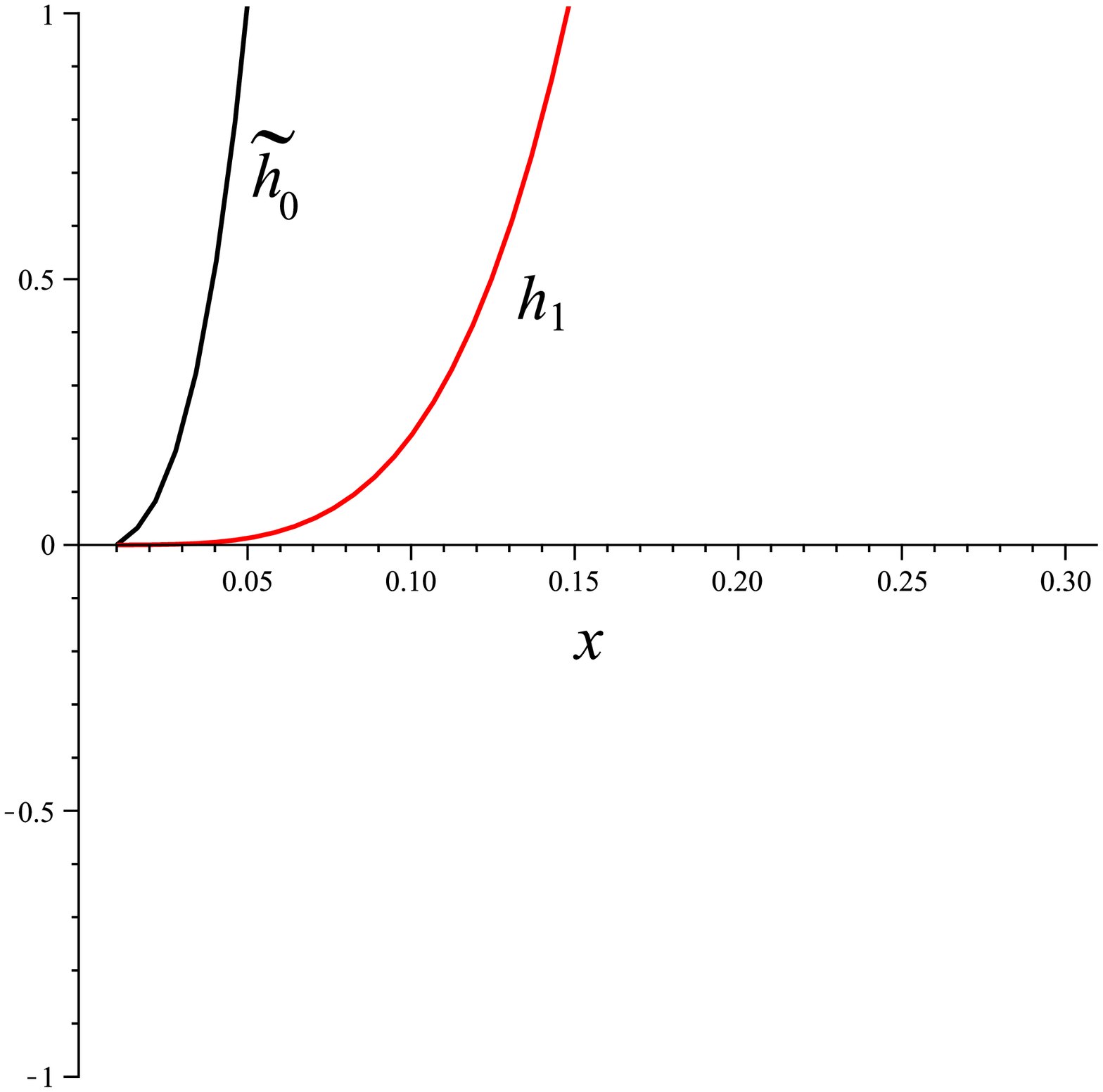}&\quad
\includegraphics[scale=0.3]{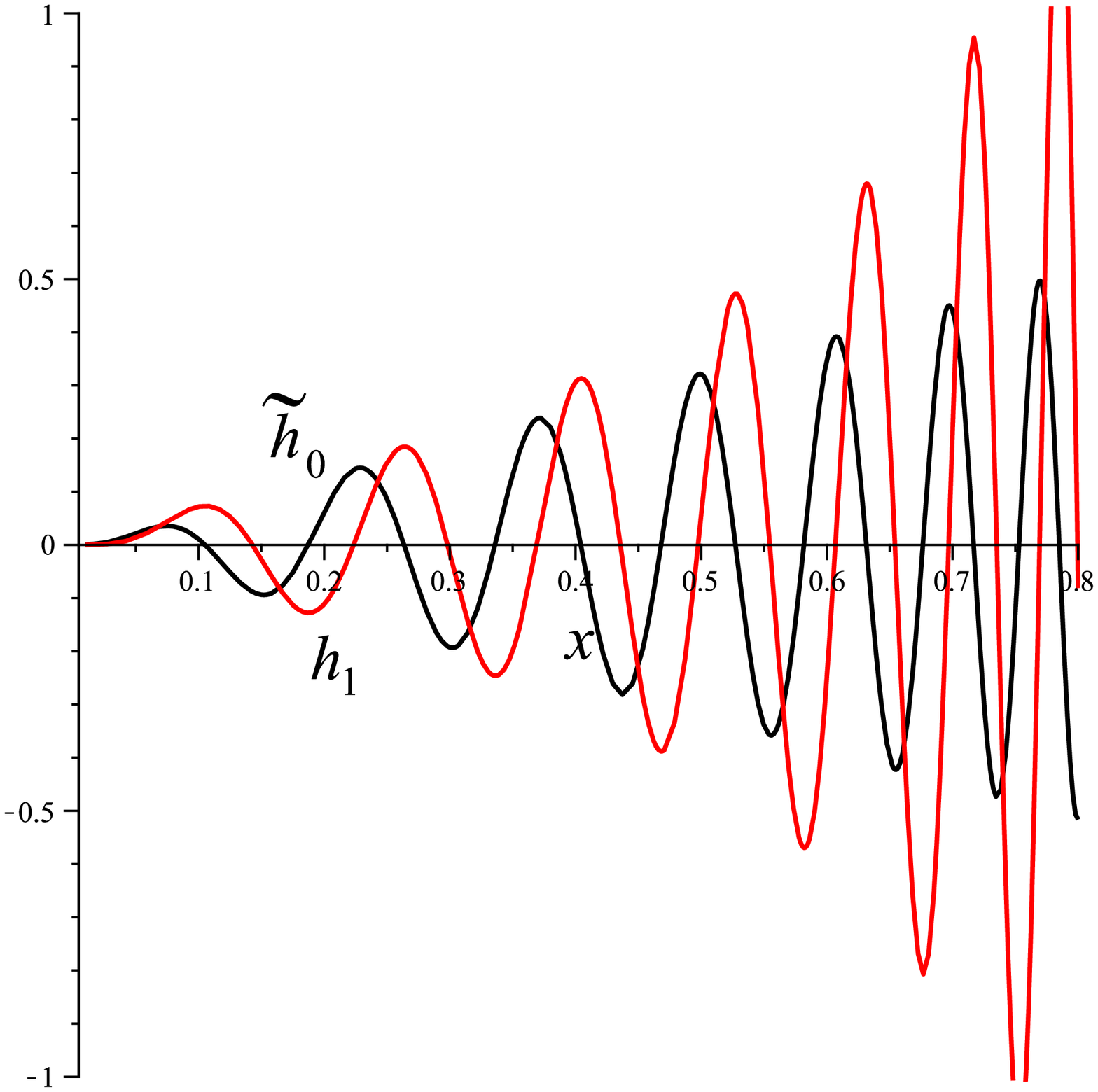}\\[.4cm]
\quad\mbox{(a)}\quad &\quad \mbox{(b)}
\end{array}$\\
\end{center}
\caption{
The behaviour of the metric perturbation functions $\tilde h_0$ 
and $h_1$ in the case of magnetic multipoles is shown in terms of 
the rescaled radial variable $x$ for $l=2$ and different values of $\Omega$.
The initial conditions are chosen so that $\tilde h_0(x_0)=10^{-4}
=h_1(x_0)$, with $x_0=10^{-2}$.
Figures (a) and (b) correspond to $\Omega=1$ and $\Omega=40$, respectively.
}
\label{fig:2}
\end{figure}

A real solution for the metric can be obtained by considering 
$\tilde h_0$ and $h_1$  as real. In this case the non-vanishing metric 
components are
\begin{eqnarray}
\fl\quad
g_{tt}&=& N^2 \,,\quad
g_{t\phi}=\tilde h_0\sin\omega t 
\sin\theta\frac{dY_{l0}}{d \theta } ,\quad
g_{r\phi}=h_1 \cos\omega t
\sin\theta\frac{dY_{l0}}{d \theta } ,
\nonumber \\
\fl\quad
g_{rr}&=& -N^{-2} ,\quad 
g_{\theta\theta}=-r^2 ,\quad
g_{\phi\phi}=-r^2\sin^2\theta  ,
\end{eqnarray}
so that
\begin{eqnarray}
\fl\quad
ds^2&=&ds_{\rm (dS)}^2 +2\sin\theta\frac{dY_{l0}}{d \theta }
\left(\tilde h_0\sin\omega t d\phi 
+2h_1 \cos\omega t dr\right) dt ,
\end{eqnarray}
to first order in the perturbation quantities.
A natural orthonormal frame associated with this form of the metric is then
\begin{eqnarray}
\Omega^{\hat 0}&=& Ndt +N^{-1}\sin\theta\frac{dY_{l0}}
{d \theta }\tilde h_0\sin\omega t d\phi, 
\nonumber \\
\Omega^{\hat r}&=& -N^{-1}dr+N\sin\theta\frac{dY_{l0}}
{d \theta }h_1 \cos\omega t d\phi,
\nonumber \\
\Omega^{\hat \theta}&=&-r d\theta ,\quad 
\Omega^{\hat \phi}=-r\sin \theta  d\phi .
\end{eqnarray}
One can then introduce a Newman--Penrose frame 
in a standard way, as in Eq. 
(\ref{NPframe}). We find that the non-vanishing first-order spin 
coefficients, Weyl scalars and Ricci terms are as follows:
\begin{eqnarray}
\fl\quad
\rho&=&-\frac{1}{r}-{\rm i}\frac{L}{2N^2r^2}[N^2h_1C+\tilde h_0S]Y_{l0},
\nonumber\\
\fl\quad
\mu&=&-\frac{N^2}{2r}+{\rm i}\frac{L}{4r^2}[N^2h_1C-\tilde h_0S]Y_{l0},
\nonumber\\
\fl\quad
\alpha&=&-\frac{\sqrt{2}}{4r}\cot\theta-{\rm i}\frac{\sqrt{2}}
{4r^2}\left[(1-2N^2)h_1C+\frac{2\omega r\tilde h_0+N^4(L-2)h_1}
{2\omega rN^2}S\right]\frac{dY_{l0}}{d\theta},
\nonumber\\
\fl\quad
\beta&=&-\alpha+{\rm i}\frac{\sqrt{2}N^2}{2r^2}h_1\frac{dY_{l0}}
{d\theta}C,
\nonumber\\
\fl\quad
\gamma&=&-\frac{H^2r}{2}+{\rm i}\frac{L}{8r^2}[N^2h_1C-\tilde h_0S]
Y_{l0},
\nonumber\\
\fl\quad
\nu&=& {\rm i}\frac{\sqrt{2}}{4r}\left\{
[N^2H^2rh_1+\omega\tilde h_0]C\right.
\nonumber\\
\fl\quad
&&\left.-\frac1{2\omega r^2}[2\omega r\tilde h_0
+N^2(N^2(L-2)-2\omega^2 r^2)h_1]S
\right\}\frac{dY_{l0}}{d\theta},
\nonumber\\
\fl\quad
\epsilon&=&-{\rm i}\frac{L}{4N^2r^2}[N^2h_1C+\tilde h_0S]Y_{l0},
\nonumber\\
\fl\quad
\kappa&=&{\rm i}\frac{\sqrt{2}}{N^4r}\left\{
[N^2H^2rh_1+\omega\tilde h_0]C\right.
\nonumber\\
\fl\quad
&&\left.+\frac1{2\omega r^2}[2\omega r\tilde h_0
+N^2(N^2(L-2)-2\omega^2 r^2)h_1]S
\right\}\frac{dY_{l0}}{d\theta},
\end{eqnarray}
and
\begin{eqnarray}
\label{psimagn}
\psi_0&=& {\rm i}\frac{1}{r^2N^4}\left\{
[N^2H^2rh_1+\omega\tilde h_0]C\right.
\nonumber\\
&&\left.+\frac1{2\omega r^2}[2\omega r\tilde h_0
+N^2(N^2(L-2)-2\omega^2 r^2)h_1]S
\right\}
{}_2Y^0{}_l,
\nonumber\\
\psi_1&=& {\rm i}\frac{\sqrt{2}(L-2)}{4r^3}\left[h_1C
-\frac1{\omega rN^2}[N^4h_1-\omega r\tilde h_0]S\right]
\frac{dY_{l0}}{d\theta},
\nonumber\\
\psi_2&=&-{\rm i} \frac{N^2}{4\omega r^4}L(L-2)h_1Y_{l0}S,
\nonumber\\
\psi_3&=& {\rm i} \frac{\sqrt{2}N^2(L-2)}{8r^3}\left[h_1C
+\frac1{\omega rN^2}[N^4h_1-\omega r\tilde h_0]S\right]
\frac{dY_{l0}}{d\theta},
\nonumber\\
\psi_4&=&-{\rm i} \frac1{4r^2}\left\{
[N^2H^2rh_1+\omega\tilde h_0]C\right.
\nonumber\\
&&\left.-\frac1{2\omega r^2}[2\omega r\tilde h_0
+N^2(N^2(L-2)-2\omega^2 r^2)h_1]S
\right\}
{}_2Y^0{}_l.
\end{eqnarray}

\begin{figure} 
\typeout{*** EPS figure S_magn}
\begin{center}
\includegraphics[scale=0.3]{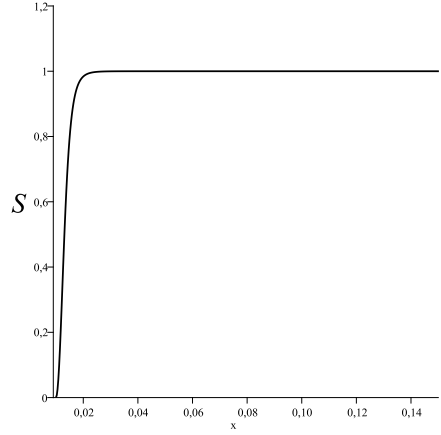}
\end{center}
\caption{The behaviour of the spectral index $\mathcal{S}$ in the case of magnetic multipoles is shown in terms 
of the rescaled radial variable $x$ for fixed values of $t=\pi/4$, $\theta=\pi/2$, $l=2$ and $\Omega=1$. 
The choice of initial conditions is the same as in Fig. \ref{fig:2} (a).
The variation of radial coordinate is limited to the range where the perturbative regime holds. 
}
\label{fig:S_magn}
\end{figure}

The behaviour of the speciality index (\ref{specdef}) as a function of the rescaled radial coordinate $x$ for fixed values of $t$ and $\theta$ is shown in Fig. \ref{fig:S_magn}. 
Note that $\mathcal{S}$ is purely real also in this case.


\section{Perturbations of radial geodesics}

We are now in a position to consider the perturbations 
\beq
\tilde U=U_{\rm geo}+u
\eeq
to the radial geodesic (\ref{rad_geo}).
We are interested in finding the solution for $u$ such that the 
path is geodesic in the perturbed gravitational field, i.e. 
$\nabla_{\tilde U}\tilde U=0$, to first order in the perturbation.

\subsection{Electric multipoles}

The perturbed 4-velocity is given by
\beq
u={\rm e}^{-{\rm i}\omega t}
\left[-\frac{\Gamma_1}{N^2}Y_{l0}\partial_t
+\zeta_1N^2Y_{l0}\partial_r+\frac{\xi_1}{r^2}\frac{d Y_{l0}}
{d\theta}\partial_\theta\right] ,
\eeq
where $\Gamma_1$, $\zeta_1$ and $\xi_1$ are functions of $r$.
Real solutions for $u$ require
\begin{eqnarray}
\label{elefunctu}
{\rm e}^{-{\rm i}\omega t}\xi_{1} 
&=&A_1\cos\omega t+A_2\sin\omega t,
\nonumber\\
{\rm e}^{-{\rm i}\omega t}\zeta_1
&=&\frac1{E\zeta_0}\left[\left(B_1
+\frac{N^2}{2}W\right)\cos\omega t+B_2\sin\omega t\right],
\nonumber\\
{\rm e}^{-{\rm i}\omega t}\Gamma_1
&=&-\frac1{E}\left[\left(B_1
+E^2W\right)\cos\omega t+\left(B_2+E\Gamma_0\zeta_0\tilde H_1\right)
\sin\omega t\right]
\end{eqnarray}
where the functions $A_1,A_2$ and $B_1,B_2$ satisfy the  
systems of equations
\begin{eqnarray}
\frac{dA_1}{dr}&=&-\frac{\omega}{\zeta_0}\left(A_2
-\frac{2E^{2}-N^{2}}{2E\omega}W\right),
\nonumber\\
\frac{dA_2}{dr}&=&\frac{\omega}{\zeta_0}\left(A_1
+\frac{\Gamma_0\zeta_0}{\omega}\tilde H_1\right),
\end{eqnarray}
and
\begin{eqnarray}
\frac{dB_1}{dr}&=&-\frac{\omega}{\zeta_0}\left(B_2
+E\Gamma_0\zeta_0\tilde H_1\right),
\nonumber\\
\frac{dB_2}{dr}&=&\frac{\omega}{\zeta_0}\left(B_1
+\frac{2E^2-N^2}{2}W\right),
\end{eqnarray}
respectively, which can be integrated numerically. 

\begin{figure} 
\typeout{*** EPS figure u_ele}
\begin{center}
$\begin{array}{cc}
\includegraphics[scale=0.3]{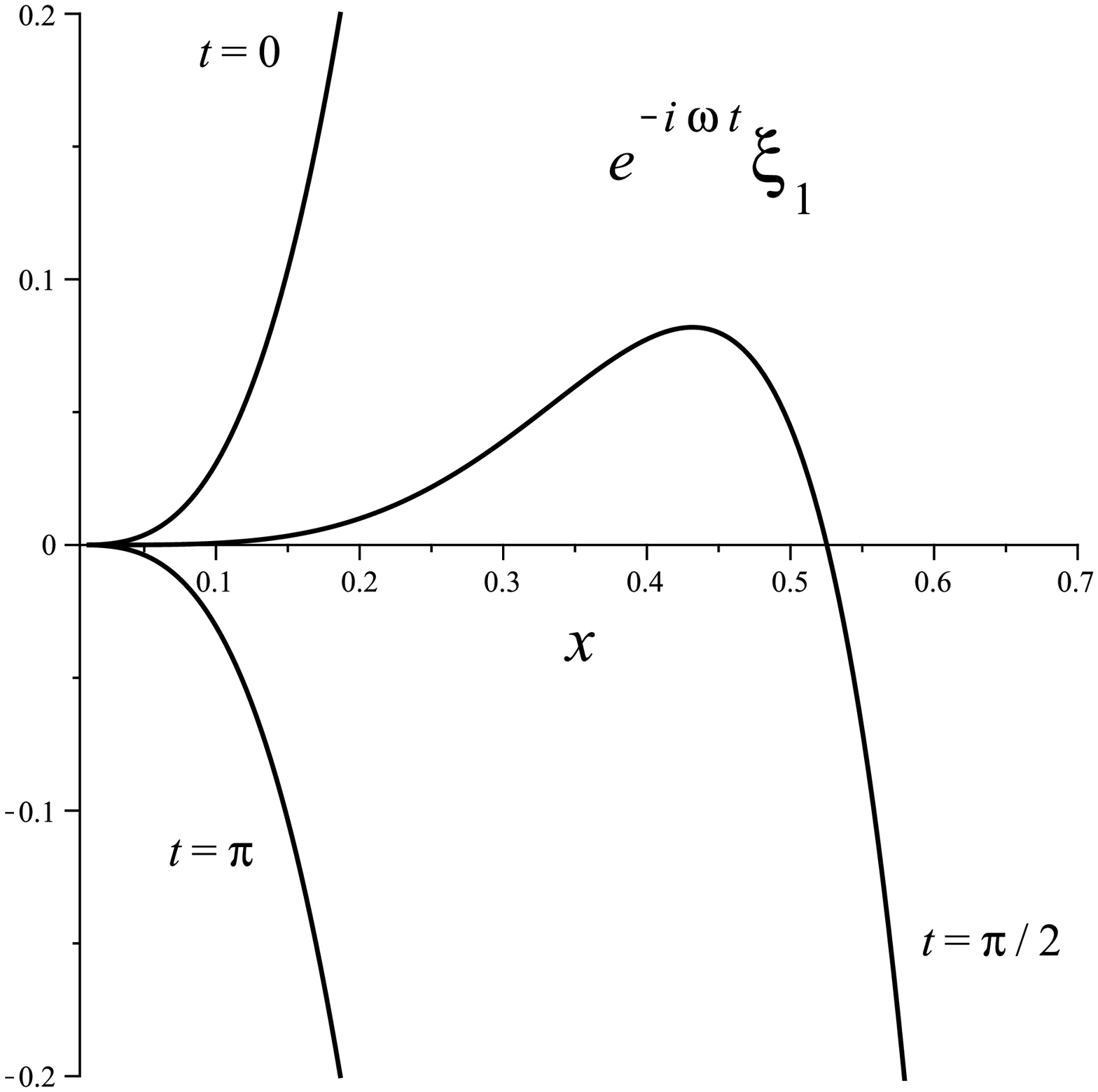}&\quad
\includegraphics[scale=0.3]{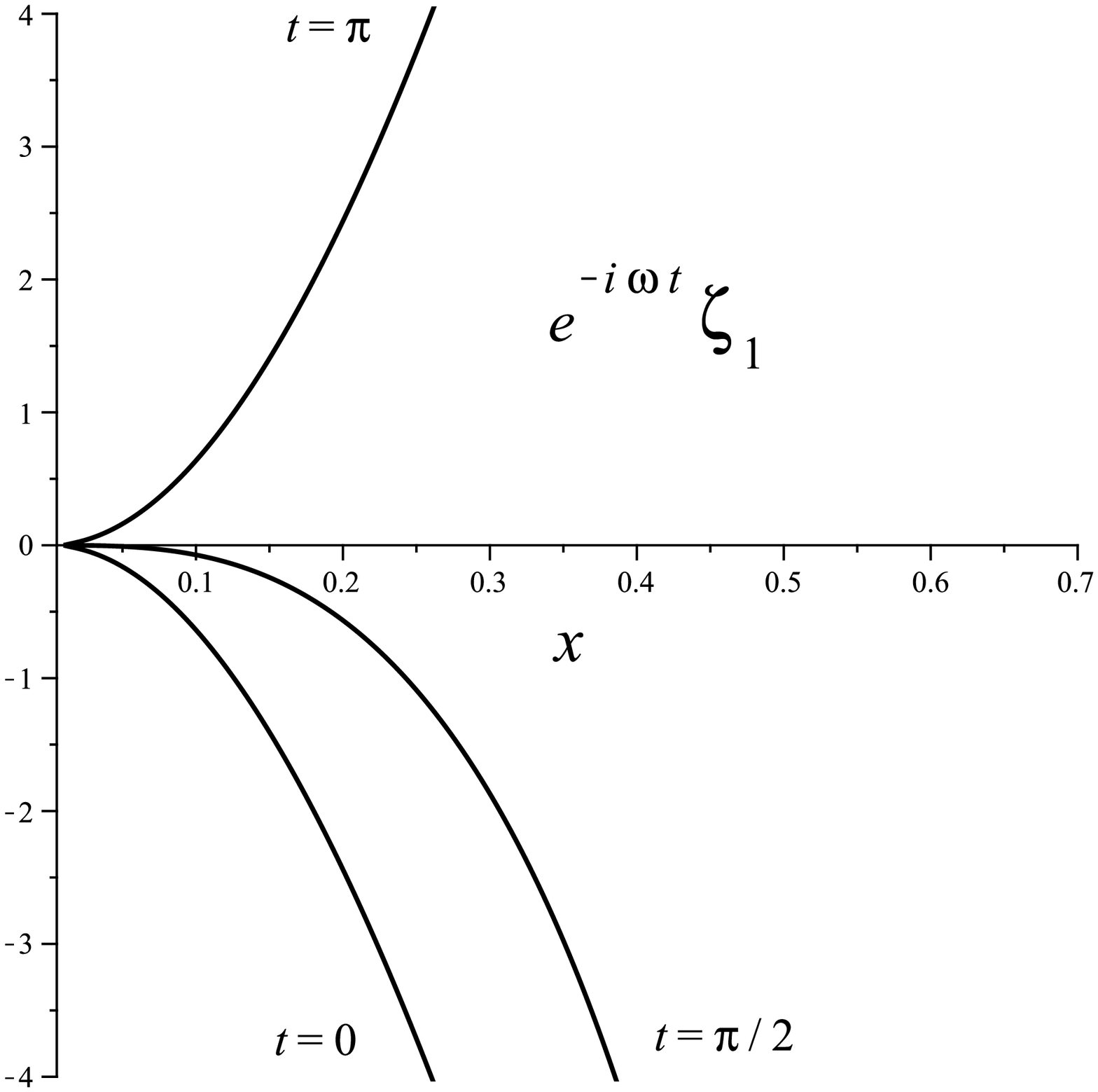}\\[.4cm]
\quad\mbox{(a)}\quad &\quad \mbox{(b)}\\[.6cm]
\end{array}$\\
\includegraphics[scale=0.3]{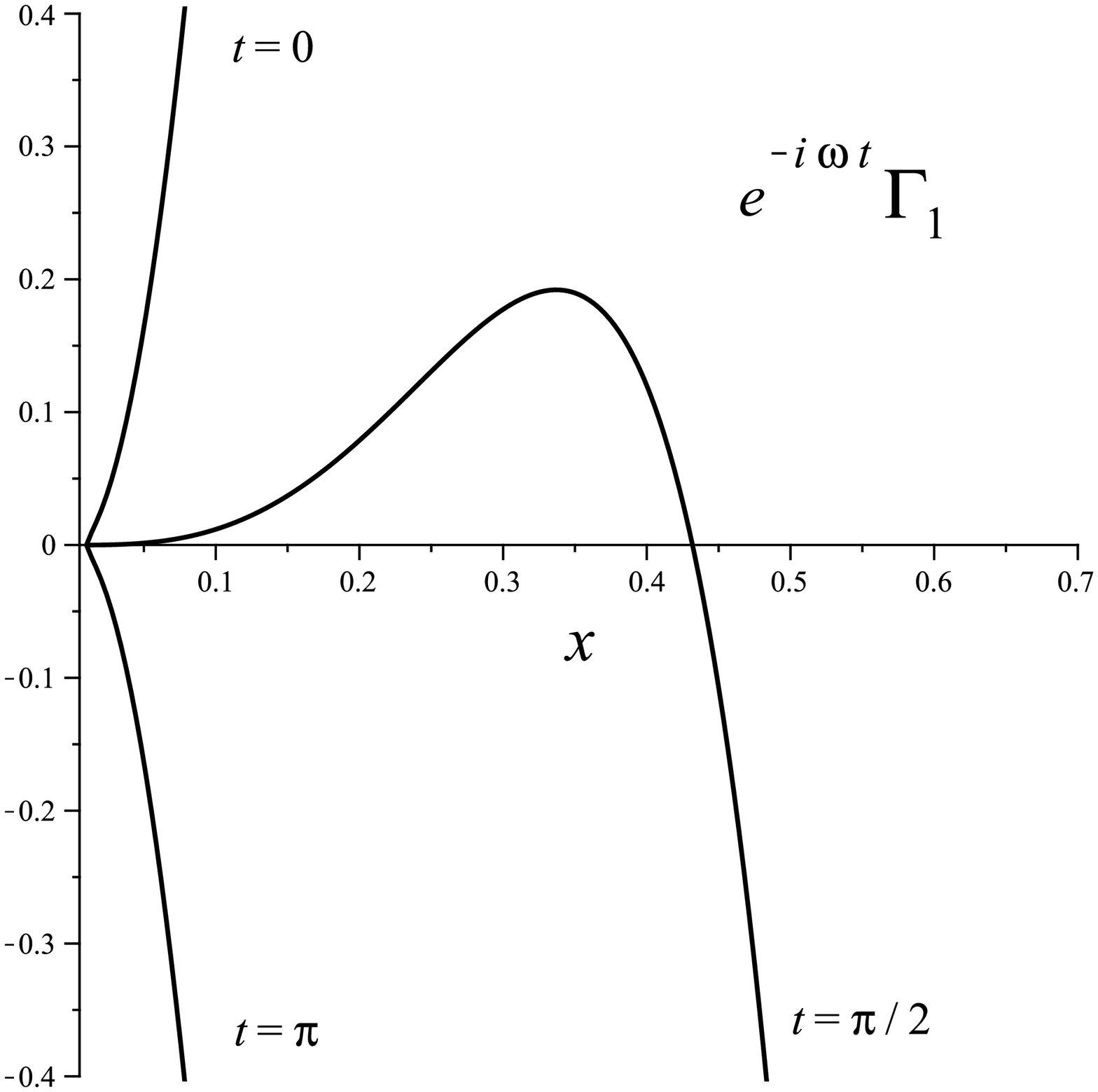}\\[.4cm]
\mbox{(c)}\\[.6cm]
\end{center}
\caption{The behavior of the perturbed velocity functions ${\rm e}^{-{\rm i}\omega t}\xi_1$, ${\rm e}^{-{\rm i}\omega t}\zeta_1$ and ${\rm e}^{-{\rm i}\omega t}\Gamma_1$ in the case of electric multipoles is shown in terms of the rescaled radial variable $x$ for different values of $t$ and the same choice of parameters as well as initial conditions for the perturbed metric functions as in Fig. \ref{fig:1} (a).
The functions $A_1,A_2$ and $B_1,B_2$ are taken instead identically vanishing at the initial point.
The unperturbed geodesic radial motion is directed outwards with energy parameter $E=1.1$.
Note that in the high frequency limit the perturbed velocity functions exhibit instead an oscillating behavior similar to that of the perturbed metric functions, the amplitude of ${\rm e}^{-{\rm i}\omega t}\xi_1$ at fixed distance being about $10^2$ smaller than that associated with $\zeta_1$ and $\Gamma_1$. 
}
\label{fig:u_ele}
\end{figure}

The behavior of the perturbed velocity functions (\ref{elefunctu}) is shown in Fig. \ref{fig:u_ele} in terms of the rescaled radial variable $x$ for different values of $t$ and fixed values of $l$, $\Omega$ and $E$.


\subsection{Magnetic multipoles}

The perturbed 4-velocity is given by
\beq
u={\rm e}^{-{\rm i}\omega t}\frac{\xi_1}{r^2\sin\theta}
\frac{d Y_{l0}}
{d \theta}\partial_\phi,
\eeq
where $\xi_1$ is a function of $r$.
Real solutions for $u$ require
\beq
\label{magnfunctu}
{\rm e}^{-{\rm i}\omega t}\xi_1=A_1\cos\omega t+A_2\sin\omega t,
\eeq
where the functions $A_1,A_2$ satisfy the system of equations
\begin{eqnarray}
\frac{dA_1}{dr}&=&-\frac{\omega}{\zeta_0}\left[A_2
+\frac{2E^{2}-N^{2}}{EN^2}\left(\tilde h_0+\frac{rH^2N^2}{\omega}
h_1\right)\right],\nonumber\\
\frac{dA_2}{dr}&=&\frac{\omega}{\zeta_0}\left\{A_1
-\frac{2\Gamma_0\zeta_0}{\omega rN^2}\left[\tilde h_0
-\omega rN^2\left(1-\frac{N^2(L-2)}{2\omega^2r^2}\right)
h_1\right]\right\},
\end{eqnarray}
which can be integrated numerically.

\begin{figure} 
\typeout{*** EPS figure u_magn}
\begin{center}
\includegraphics[scale=0.3]{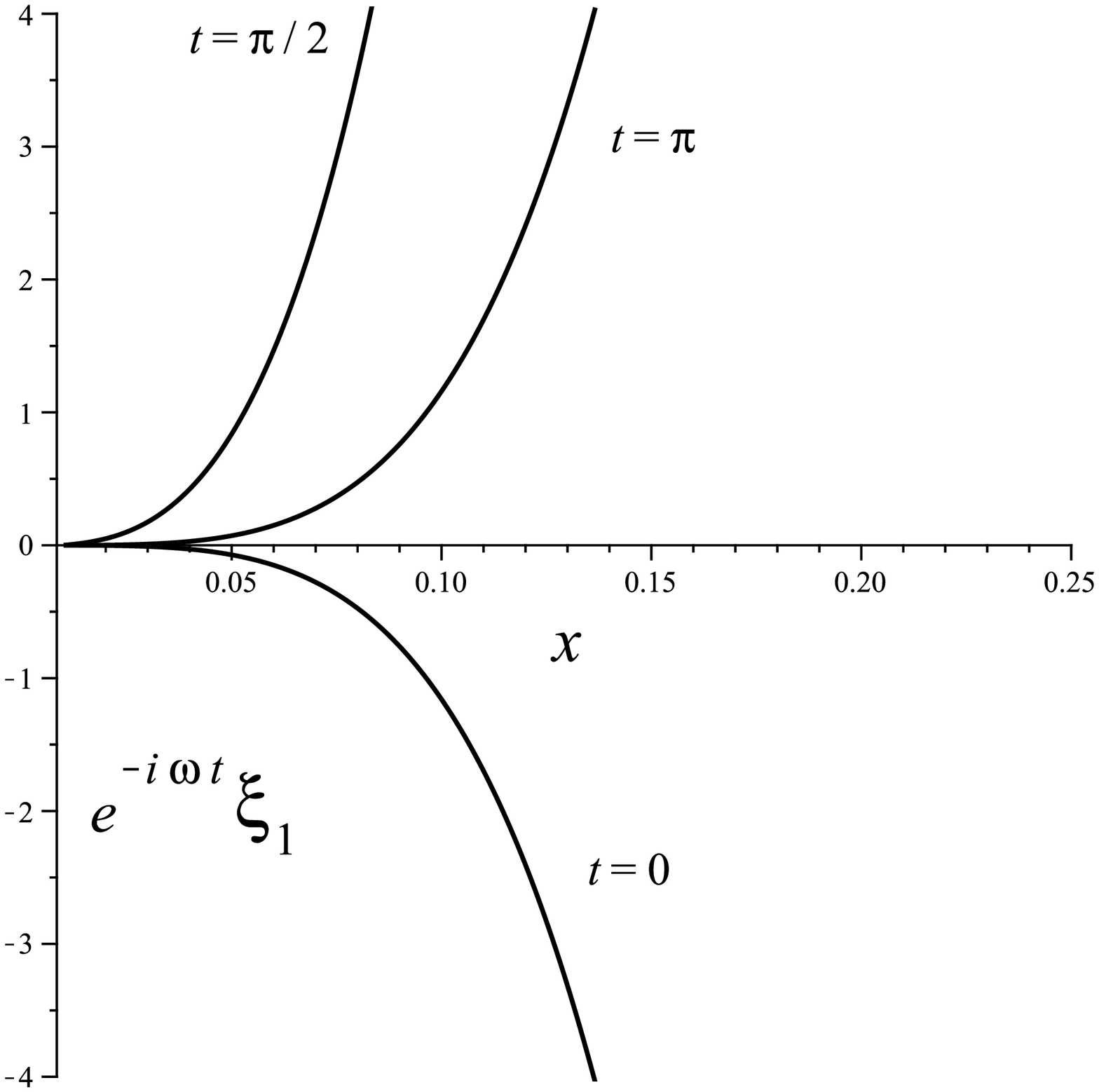}
\end{center}
\caption{The behavior of the perturbed velocity function ${\rm e}^{-{\rm i}\omega t}\xi_1$ in the case of magnetic multipoles is shown in terms of the rescaled radial variable $x$ for different values of $t$ and the same choice of parameters as well as initial conditions for the perturbed metric functions as in Fig. \ref{fig:2} (a).
The functions $A_1,A_2$ are taken instead identically vanishing at the initial point.
The unperturbed geodesic radial motion is directed outwards with energy parameter $E=1.1$.
Note that in the high frequency limit the perturbed velocity function exhibits instead an oscillating behavior similar to that of the perturbed metric functions.
}
\label{fig:u_magn}
\end{figure}

The behavior of the perturbed velocity function (\ref{magnfunctu}) is shown in Fig. \ref{fig:u_magn} in terms of the rescaled radial variable $x$ for different values of $t$ and fixed values of $l$, $\Omega$ and $E$.


\section{Teukolsky equation}

In this section we study gauge- and tetrad-invariant first-order 
massless perturbations of any spin ($s= \pm2$ for gravitational waves, 
$s= \pm1$ for electromagnetic waves, and $s= \pm1/2$ for neutrinos) 
in the de Sitter background following the approach of Teukolsky \cite{Teuk}.
Although this problem have been already addressed in the literature (see, e.g., Ref. \cite{galtsov}, where the perturbation equations for fields of different spin have been solved by using the technique of Debye potentials), we are are interested here in investigating the existence of closed-form, i.e. Liouvillian, solutions to the perturbation equations.

The Teukolsky master equation, in this case, reads as
\begin{eqnarray}
\label{teukgen}
\fl\qquad
0&=&\biggr \{ \frac{1}{2N^2}\partial_{tt}-\frac{N^2}{2}\partial_{rr}
-\frac{1}{2r^2}\partial_{\theta\theta}-\frac{1}{2r^2\sin^2\theta}
\partial_{\phi\phi}
\nonumber\\
\fl\qquad
&&+\frac{s}{rN^2}\partial_t+\frac{(s+1)(1-2N^2)}{r}\partial_r
-\frac{\cot\theta}{2r^2}\partial_\theta 
\nonumber\\
\fl\qquad
&&-\frac{is\cos\theta}{r^2\sin^2\theta}\partial_{\phi}
+\frac{1}{2r^2}\left[(1+s)[3s+2-2N^2(2s+1)]
+\frac{s^2}{\sin^2\theta}\right]\biggr \} \psi.
\end{eqnarray}
Note that this equation is solved by $\psi=\psi_0$ with $s=2$ and 
by $\psi=\psi_4/\rho^4$ with $s=-2$, where $\psi_0$ and $\psi_4$ are 
the perturbed Weyl scalars given by Eqs. (\ref{psiele}) and 
(\ref{psimagn}) for electric and magnetic multipoles, respectively.   
Therefore, the latter equations provide a link between the perturbed 
metric functions and the solutions to the Teukolsky equation.

Let us now study the Teukolsky equation in completely general form, 
relaxing the assumption of azimuthal symmetry.
The master equation (\ref{teukgen}) admits separable solutions of the form
\beq
\psi(t,r,\theta,\phi)
={\rm e}^{-{\rm i}\omega t}{\rm e}^{{\rm i} m \phi} R(r)S(\theta) ,
\eeq
where $\omega>0$ is the wave frequency and $m$ is the azimuthal separation 
constant.

The angular function satisfies the equation of the spin-weighted 
spherical harmonics, i.e.
\beq
\frac{1}{\sin\theta}\frac{\rmd }{\rmd \theta} 
  \left(\sin\theta \frac{\rmd S(\theta)}{\rmd \theta}\right)
  + V_{\rm (ang)}(\theta)S(\theta)=0 ,
\eeq
where
\begin{eqnarray}
V_{\rm (ang)}(\theta)
& = &L-s^2-\frac{(s\cos\theta+m)^2}{\sin^2\theta} ,
\end{eqnarray}
and the separation constant has been chosen according to Teukolsky.

The radial equation can be cast in the form 
\beq
\label{radeq}
Q^{-s} \frac{\rmd}{\rmd r}\left(Q^{s+1}\frac{\rmd R(r)}{\rmd r} \right)
+V_{\rm (rad)}(r)R(r)=0 ,
\eeq
where $Q = r^{2} N^{2}$ and
\begin{eqnarray}\fl
\label{radiale1}
V_{\rm (rad)}(r)& = &
\frac{1}{N^4}\left\{\omega^2+\frac{2{\rm i}\omega s}{r}
-\frac{N^2}{r^2}[L+(1+s)(3s+2)]+2\frac{N^4}{r^2}(1+s)(2s+1)\right\}.
\end{eqnarray}

\subsection{Liouvillian solutions to the radial equation}

The general solution to the radial part of the Teukolsky equation can 
be expressed in terms of Heun functions.
We are interested here in investigating the existence of closed-form, 
i.e. Liouvillian, solutions to the radial equation (\ref{radeq}) governing 
the perturbations. Roughly speaking, the set of Liouvillian functions 
includes the usual elementary functions such as
exponential, trigonometric, logarithmic functions, etc., but not generic 
hypergeometric functions or other special functions.
We apply the well-known Kovacic algorithm \cite{kovacic} 
outlined in appendix D.

The radial equation (\ref{radeq}) can be converted in normal form, i.e. 
\beq
\label{radeqNF}
\frac{\rmd^2y(r)}{\rmd r^2}=W_{\rm (rad)}(r)y(r),
\eeq
by the scaling
\beq
R = yQ^{-(1+s)/2} ,
\eeq
where the function $W_{\rm (rad)}$ is given by
\beq\fl\qquad
\label{Wdef}
W_{\rm (rad)}(r) = \frac{c_0}{r^2}+\frac{c_H^-}{(r-r_H)^2}
+\frac{c_H^+}{(r+r_H)^2}+\frac{d_0}{r}+\frac{d_H^-}{(r-r_H)}
+\frac{d_H^+}{(r+r_H)},
\eeq
with parameters
\begin{eqnarray}
\fl\qquad
c_0& = &L, \qquad
c_H^{\pm} = \frac{1}{4}[(s\pm {\rm i}\omega r_H)^2-1], 
\nonumber\\
\fl\qquad
d_0& = &-2{\rm i}\omega s, \qquad
d_H^{\pm} = \mp
\frac{1}{4r_H}[\omega^2r_H^2\mp4{\rm i}\omega r_Hs-s^2-2L+1].	
\end{eqnarray}

Let us discuss in detail the case of type 1 solutions 
(see again appendix D). For all possible 
pole structures of (\ref{Wdef}), all functions 
$\tilde\Omega$ generated by the Kovacic algorithm have the form
\beq
\tilde\Omega=\frac{\alpha_0}{r}+\frac{\alpha_{H}^+}{(r+r_H)}
+\frac{\alpha_{H}^-}{(r-r_H)},
\eeq
where
\beq
\alpha_{0}=\frac12\left[1+\delta_0\sqrt{1+4c_0}\right] , \qquad
\alpha_H^{\pm}=\frac12\left[1+\delta_H^\pm\sqrt{1+4c_H^\pm}\right] ,
\eeq
the quantities $\delta_0$, $\delta_H^+$ and $\delta_H^-$ taking 
independently values $\pm1$.

The necessary conditions for type 1 solutions are fulfilled if the 
following constraint is imposed:
\beq
\label{cond1}
d_0+d_H^++d_H^-=0 ,
\eeq
implying that the order of the pole at infinity is $2$.

The algorithmic condition $d=n$ is given by 
\beq
\label{cond2}
\alpha_\infty-\alpha_0-\alpha_{H}^+-\alpha_{H}^-=n ,
\eeq
where
\beq
\alpha_{\infty}=\frac12\left[1+\delta_\infty\sqrt{1+4(c_0+c_H^+
+c_H^-)}\right]\ , \qquad \delta_\infty=\pm1 .
\eeq

For every pole structure, we construct the Liouvillian solutions 
to Eq. (\ref{radeqNF}) by deriving a consistent
recursion relation which generates the finite set of coefficients 
in the polynomial ${\mathcal P}$. 
To obtain a recursion relation with the fewest number of terms we 
take, without loss of generality, ${\mathcal P}$ to be expanded about 
a pole of non-vanishing order, say $r=0$, as follows:
\beq
{\mathcal P}=\sum_{k=0}^n a_kr^k\ , \qquad a_n\not=0 .
\eeq
After substituting this expression for ${\mathcal P}$ into Eq. 
(\ref{eqPdir}) we obtain in the general case, when all three poles 
are present, the recursion relation
\begin{eqnarray}
\label{recureq}
0&=&\Bigr[(k-1)(k-2-2n+2\alpha_\infty)-r_H(g_H^+-g_H^-)\Bigr]a_{k-1}
\nonumber\\
&&-r_{H}\Bigr[2k(\alpha_H^+-\alpha_H^-)-r_H(g_H^++g_H^-)\Bigr]a_{k}
\nonumber\\
&&-(k+1)r_H^2(k+2\alpha_0)a_{k+1} , \qquad k=0,\ldots,n+1,
\end{eqnarray}
where 
\begin{eqnarray}
g_0& = &-d_0+2\frac{\alpha_0}{r_H}
(\alpha_H^+-\alpha_H^-) , \nonumber\\
g_H^\pm& = &-d_H^\pm\mp
\frac{\alpha_H^\pm}{r_H}(2\alpha_0+\alpha_H^\mp),
\end{eqnarray}
with the understanding that $a_i=0$ if $i\leq-1$ or $i\geq n+1$.

Liouvillian solutions (\ref{ysol}) to Eq. (\ref{radeqNF}) are thus given by
\beq
y={\mathcal P}r^{\alpha_0}(r+r_H)^{\alpha_H^+}(r_H-r)^{\alpha_H^-} ,
\eeq
implying that 
\beq
R={\mathcal P}r_{H}^{1+s}    
r^{\alpha_0-\frac{1+s}2}(r+r_H)^{\alpha_H^+
-\frac{1+s}2}(r_H-r)^{\alpha_H^--\frac{1+s}2} .
\eeq

\section{Concluding remarks}

The present work has been motivated by a previous study on gravitational waves
about de Sitter backgrounds \cite{Bini2008}, which is relevant for
the stochastic background emerging from inflation, and by the analysis
in \cite{Ferrari1988} of gravitational waves and their geodesics.
We have solved the  gravitational perturbation problem in the static region of the de Sitter spacetime between the origin and the cosmological horizon by using the Regge--Wheeler formalism.
The set of perturbation equations has been reduced to a single second order differential equation of the Heun-type for both electric and magnetic multipoles.
We have then investigated the geodesic motion in the perturbed gravitational field.
In particular, an initially radial geodesic in the unperturbed spacetime  cannot remain radial as a result of the perturbation. 
The deviation from radial motion has been shown to occur in both polar and azimuthal directions. 
Finally, we have investigated gauge- and tetrad-invariant first-order massless perturbations of any spin following the approach of Teukolsky.
The general solution to the radial part of the Teukolsky equation can be expressed in terms of Heun functions.
However, we have shown that there also exists in addition a class of closed-form, i.e. Liouvillian, solutions.

\appendix

\section{Electric multipoles: $l=0,1$}

\subsection{$l=0$}

The system of perturbation equations in the case $l=0$ is given by
\begin{eqnarray}
\label{eqseleleq0}  
0&=&H_0{}'-\frac{2\omega}{N^2}\tilde H_1-\frac{rH^2}{N^2}H_2
-\frac{r}{N^4}(\omega^2+H^2)K , 
\nonumber\\
0&=&K{}'-\frac{H_2}{r}+\frac{K}{rN^2} . 
\end{eqnarray}
We can set  
\beq
H_0=H_2\equiv W, \qquad \tilde H_1=0,
\eeq
without loss of generality, so that Eq. (\ref{eqseleleq0}) reduces to
\begin{eqnarray}
\label{eqseleleq0_new}  
0&=&W{}'-\frac{rH^2}{N^2}W-\frac{r}{N^4}(\omega^2+H^2)K , 
\nonumber\\
0&=&K{}'-\frac{W}{r}+\frac{K}{rN^2} . 
\end{eqnarray}
In terms of the dimensionless variable $x \equiv Hr$ 
the corresponding solution is
\begin{eqnarray}\fl\qquad
K&=&\frac{\sqrt{1-x^2}}{x}\left[c_1\left(\frac{1-x}{1+x}\right)^{
\sqrt{1+\Omega^2}/2}+c_2\left(\frac{1-x}{1+x}\right)^{-\sqrt{1
+\Omega^2}/2}\right] , 
\nonumber\\
\fl\qquad
W&=&\frac{\sqrt{1+\Omega^2}}{\sqrt{1-x^2}}\left[-c_1\left(
\frac{1-x}{1+x}\right)^{\sqrt{1+\Omega^2}/2}+c_2\left(\frac{1-x}{1+x}
\right)^{-\sqrt{1+\Omega^2}/2}\right] 
\end{eqnarray}
where $\Omega \equiv \omega/H$.

\subsection{$l=1$}

The system of perturbation equations in the case $l=1$ is given by
\begin{eqnarray}
\label{eqseleleq1}  
0&=&H_0{}'-\frac{H_0}{rN^2}+\frac{\omega^2r}{N^4}K
-\frac{rH^2}{N^2}H_{2} , 
\nonumber\\
0&=&K{}'-\frac{H_2}{r} , 
\end{eqnarray}
plus the algebraic relation
\beq
\tilde H_1=\frac{\omega r}{N^2}K.
\eeq
We can set  
\beq
H_0=H_2\equiv W,
\eeq
so that Eq. (\ref{eqseleleq1}) becomes
\begin{eqnarray}
\label{eqseleleq1_new}  
0&=&W{}'+\frac{\omega^2r}{N^4}K-\frac{(1+H^{2}r^{2})}{rN^2}W , 
\nonumber\\
0&=&K{}'-\frac{W}{r} . 
\end{eqnarray}
In terms of the dimensionless variable $x \equiv Hr$ 
the corresponding solution is
\begin{eqnarray}
K&=&c_1\sin(\Omega\,{\rm arctanh}\,x)+c_2\cos(\Omega\,
{\rm arctanh}\,x) , \nonumber\\
W&=&\frac{\Omega x}{(1-x^{2})}\left[c_1\cos(\Omega\,{\rm arctanh}\,x)
-c_2\sin(\Omega\,{\rm arctanh}\,x)\right] . 
\end{eqnarray}

\section{Magnetic multipoles: $l=1$}

The system of perturbation equations in the case $l=1$ reduces to 
\beq
0=\tilde h_0{}'-\frac{2\tilde h_0}{r}+\omega h_{1}.
\eeq
Without loss of generality we can set $h_1=0$, so that the previous 
equation becomes
\beq
0=\tilde h_0{}'-\frac{2\tilde h_0}{r},
\eeq
whose solution is simply
\beq
\tilde h_0=c_1r^{2}.
\eeq

\section{Decoupling the perturbation equations}

The system of perturbation equations of both parity can be reduced to a 
second-order differential equation for a single perturbation function, 
the remaining ones being determined either algebraically or by differentiation.
Such a master equation can be cast in general in the form of a Heun 
equation \cite{heun,ronveaux}
\beq
\label{heun}
\frac{d^{2}y}{dz^2}+\left(\frac{\gamma}{z}
+\frac{\delta}{z-1}+\frac{\epsilon}{z-a}\right)\frac{dy}{dz}
+\frac{\alpha\beta-q}{z(z-1)(z-a)}y=0,
\eeq
where $\{\alpha,\beta,\gamma,\delta,\epsilon,a,q\}$ are complex 
arbitrary parameters such that $\gamma+\delta+\epsilon
=\alpha+\beta+1$ and $a\not=0,1$.
This equation has four regular singular points, at $\{0, 1, a,\infty\}$.   

This is a generalization of the hypergeometric equation.
In fact, in the special case where $\epsilon=0$ and $q=\alpha\beta a$, 
Eq. (\ref{heun}) becomes
\beq
\label{hypergeom}
z(z-1)\frac{d^2y}{dz^2}+[(\alpha+\beta+1)
-\gamma]\frac{dy}{dz}+\alpha\beta y=0,
\eeq
which is just the hypergeometric equation in its standard form, with solution
\beq
y=c_1F(\alpha,\beta;\gamma;z)+c_2z^{1-\gamma}F(\alpha-\gamma+1,
\beta-\gamma+1;2-\gamma;z).
\eeq

\subsection{Electric multipoles}

The coupled system (\ref{eqsele2a})--(\ref{eqsele2b}) is equivalent to 
solve a second-order differential equation for $K$ of the Heun type.
The procedure is as follows. First,
solve algebraically Eq. (\ref{eqsele2b}) for $\tilde H_1$ and substitute 
it into Eq. (\ref{eqsele2a}), thus obtaining a second-order differential 
equation for $K$. Introduce then the new variable $z = x^2$ 
togheter with the rescaling 
\beq\fl\quad
\label{rescaling}
K(z) = f(z)y(z),\quad
f(z) = z^{\alpha_1}(z-1)^{\alpha_2}(z-a)^{\alpha_3},\quad
a = -\frac{L}{2}\frac{(L-2)}{(L+2\Omega^{2})}.
\eeq
The function $y$ thus satisfies an equation of the form (\ref{heun}) 
with parameters
\begin{eqnarray}
\gamma& = &2\alpha_1+\frac32,\quad
\delta = 2\alpha_2+1,\quad
\epsilon = 2\alpha_3-1,
\nonumber\\
\alpha& = &\alpha_1+\alpha_2+\alpha_3
+\frac14-\frac{\sigma_3}{4},\quad
\beta = \alpha+\frac{\sigma_3}{2},
\nonumber\\
q& = &3(\alpha_1+\alpha_2+1)(1-a)+a(\alpha\beta-2)
-3\alpha_2+\frac{L}{4},
\end{eqnarray}
and
\beq
\alpha_1 = -\frac14+\frac{\sigma_1}{4}\sqrt{1+4L}\,,\quad
\alpha_2 = -{\rm i} \sigma_2\frac{\Omega}{2},\quad
\alpha_3 = 2,
\eeq
with $\sqrt{1+4L}=2l+1$ and $\sigma_1$, $\sigma_2$, $\sigma_3$ taking 
independently the values $\pm1$.
The signs have to be chosen so as to avoid singular 
behaviours at $z=0$ and $z=1$.

\subsection{Magnetic multipoles}

The same procedure outlined above for the case of electric multipoles 
can be followed here to reduce the coupled system 
(\ref{eqsmagn2a})--(\ref{eqsmagn2b}) to a second-order differential 
equation for $h_1$ of the Heun type. First, eliminate 
$\tilde h_0$ in favour of $h_1$, then introduce the new variable 
$z = x^2$ and a rescaling $h_1(z) = f(z)y(z)$ 
of the form (\ref{rescaling}).
The function $y$ thus satisfies an equation of the form (\ref{heun}) 
with parameters
\begin{eqnarray}
\gamma& = &2\alpha_1-\frac12,\quad
\delta = 2\alpha_2+3,\quad
\epsilon = 2\alpha_3,
\nonumber\\
\alpha& = &\alpha_1+\alpha_2+\frac34-\frac{\sigma_3}{4}\,,\quad
\beta = \alpha+\frac{\sigma_3}{2},
\nonumber\\
q& = &\alpha\beta a,
\end{eqnarray}
and
\beq
\alpha_1 = \frac34+\frac{\sigma_1}{4}\sqrt{1+4L},\quad
\alpha_2 = -1-{\rm i}\sigma_2\frac{\Omega}{2},\quad
\alpha_3 = 0,
\eeq
implying that $\epsilon=0$.
Therefore in this case the Heun equation reduces to the hypergeometric 
equation (\ref{hypergeom}).  

\section{The Kovacic algorithm: an overview}

The Kovacic algorithm applies to a second-order ordinary 
differential equation of normal form 
\beq
\label{2ordNF}
\frac{\rmd^2y}{\rmd r^2}=W(r)y,
\eeq
when $W(r)$ is a given element of $\mathbb{C}(r)$, the field of rational 
functions with coefficients in the field of complex numbers.
Kovacic proved that all Liouvillian solutions to Eq. (\ref{2ordNF}) 
belong to three mutually exclusive types.
Furthermore, for each type, he provided an algorithm which 
determines whether a Liouvillian solution exists
for that type, and constructs the solution 
in case it does exist. The types are as follows:

\begin{itemize}

\item[Type 1.]
The differential equation has a solution 
of the form ${\rm e}^{\int\tilde\Omega\rmd r}$, where 
$\tilde\Omega\in\mathbb{C}(r)$.

\item[Type 2.]
The differential equation has a solution of the form 
${\rm e}^{\int\tilde\Omega\rmd r}$, where 
$\tilde\Omega$ is algebraic over $\mathbb{C}(r)$ of degree 2 
(i.e., $\tilde\Omega$ is a solution of a polynomial equation of 
degree 2 with coefficients in $\mathbb{C}(r)$), and 
solutions of type $1$ form an empty set.

\item[Type 3.]
All solutions of the differential equation are algebraic over 
$\mathbb{C}(r)$ of degree $2$, and solutions of type $1$ and $2$
form two empty sets. 

\end{itemize}

If none of the previous cases applies, the 
differential equation has no Liouvillian solution.
The construction of each type of Liouvillian solution is derived from 
the pole structure of the rational function.
Kovacic also discussed some conditions that are necessary for 
cases 1, 2 or 3 to hold. 
For type 1 every pole of $W(r)$ should have even 
order or else order 1, and the pole at $\infty$ should have either 
even order or order greater than 2.
For type 2 a necessary condition is that $W(r)$ must have at 
least one pole that has either odd order greater than 2 or order 2.
Last, for type 3 the order of a pole of $W(r)$ cannot exceed 2, 
and the order of the pole at $\infty$ must be at least 2.

Let us discuss in detail the case 1. The remaining cases 2 and 3 
apply as well and can be treated similarly.

For a fixed rational function $W(r)$, type $1$ Liouvillian solutions 
of Eq. (\ref{2ordNF}) are determined in the following way. 
A finite set of constants $d$ is derived algorithmically from the pole 
structure of $W(r)$ and for each $d$ a function $\tilde\Omega(r)$ 
is also constructed. If any $d$ is a
non-negative integer $n$ and, for the corresponding $\tilde\Omega(r)$, 
the differential equation
\beq
\label{eqPdir}
\frac{\rmd^{2}{\mathcal P}(r)}{\rmd r^2}+2\tilde\Omega\frac{\rmd {\mathcal P}(r)}
{\rmd r} + \left[\frac{\rmd\tilde\Omega}{\rmd r}
+\tilde\Omega^2-W(r)\right]{\mathcal P}(r)
=0 
\eeq
has a polynomial solution ${\mathcal P}$ of degree $n$, then Eq. 
(\ref{2ordNF}) has a Liouvillian solution given by
\beq
\label{ysol}
y={\mathcal P}{\rm e}^{\int\tilde\Omega\rmd r} .
\eeq
If this is not the case for any $d$, then Eq. (\ref{2ordNF}) 
has no type $1$ solutions.

Since $W(r)$ depends on different independent parameters, several 
different pole structures may occur by imposing conditions on the 
parameters which alter the order of poles. For each different pole 
structure, in general, $d$ and $\tilde\Omega$ are expressed in terms 
of the parameters. Solutions to Eq. (\ref{2ordNF}) may be obtained by 
requiring $d=n$, which then is just a constraint on the parameters, and 
finding all sets of parameters for which Eq. (\ref{eqPdir}) has a 
polynomial solution of degree $n$.

\section*{Acknowledgement}
ICRANet is thanked for support and Dr. V. Montaquila for stimulating 
discussion. GE is grateful to Dipartimento di Scienze Fisiche of 
Federico II University, Naples, for hospitality and support, and he
dedicates to Maria Gabriella his contribution to this work.

\end{document}